\documentclass[prl,twocolumn,superscriptaddress,showpacs,floatfix]{revtex4-1}

\usepackage{mathrsfs}
\usepackage{amssymb, amsbsy, amsmath, latexsym, dsfont, array, layout,mathrsfs,ulem,bm}
\usepackage[dvipsnames]{xcolor}
\usepackage{multirow}
\usepackage[colorlinks,linkcolor=blue,anchorcolor=red,citecolor=blue]{hyperref}
\usepackage{graphicx}
\usepackage{xspace, stmaryrd, ulem}

\definecolor{delete}{rgb}{1.0, 0.0, 0.0}
\definecolor{edit}{rgb}{0.0, 0.0, 0.9}
\definecolor{comment}{rgb}{0.9, 0.0, 0.0}

\newcommand{\one}{\mathds{1}}
\newcommand{\ket}[1]{\left|{#1}\right\rangle}
\newcommand{\bra}[1]{\left\langle{#1}\right|}
\newcommand{\braket}[2]{\langle{#1}|{#2}\rangle}

\newcommand{\ketbra}[2]{\left|{#1}\rangle\!\langle{#2}\right|}

\newcommand{\etal}{\textit{et al.}}

\newcommand{\eg}{\textit{e. g.}}
\newcommand{\de}[1]{\left(#1\right)}
\newcommand{\De}[1]{\left[#1\right]}
\newcommand{\DE}[1]{\left\{#1\right\}}
\newcommand{\mean}[1]{\left<#1\right>}
\newcommand{\abs}[1]{\left|#1\right|}

\newcommand{\y}{\mathbf{y}}
\newcommand{\bb}{\mathbf{b}}
\newcommand{\p}{\mathbf{p}}

\begin{document}

\title{
Synchronous observation of Bell nonlocality and state-dependent Kochen-Specker contextuality}

%%%%%%%%%%%%%%%%%%%%%%%%%%%%%%%%%%%%%%%%%%%%%%%%%%%%%%%%%%%%%%%%%%%

\author{Lei Xiao}%\email{xiaolei9210@126.com}
\affiliation{Beijing Computational Science Research Center, Beijing 100084, China}

\author{G. Ruffolo}
\affiliation{Instituto de F\'{i}sica ``Gleb Wataghin'', Universidade Estadual de Campinas, 130830-859, Campinas, Brazil}

\author{A. Mazzari}
\affiliation{Instituto de F\'{i}sica ``Gleb Wataghin'', Universidade Estadual de Campinas, 130830-859, Campinas, Brazil}

\author{T. Temistocles}
\affiliation{Departamento de F\'{i}sica, Instituto de Ci\^{e}ncias Exatas, Universidade Federal de Minas Gerais, 30123-970, Belo Horizonte, Brazil}
\affiliation{Instituto Federal de Alagoas - Campus Penedo, Rod. Eng. Joaquim Gon\c{c}alves - Dom Constantino, 57200-000, Penedo, AL, Brazil}

\author{M. Terra Cunha}
\affiliation{Instituto de Matem\'{a}tica, Estat\'{i}stica e Computaç\~{a}o Cient\'{i}fica, Universidade Estadual de Campinas, 130830-859, Campinas, Brazil}

\author{R. Rabelo}\email{rabelo@ifi.unicamp.br}
\affiliation{Instituto de F\'{i}sica ``Gleb Wataghin'', Universidade Estadual de Campinas, 130830-859, Campinas, Brazil}

\author{Peng Xue}\email{gnep.eux@gmail.com}
\affiliation{Beijing Computational Science Research Center, Beijing 100084, China}

%%%%%%%%%%%%%%%%%%%%%%%%%%%%%%%%%%%%%%%%%%%%%%%%%%%%%%%%%%%%%%%%%%%

\date{\today}

\begin{abstract}
Bell nonlocality and Kochen-Specker contextuality are two remarkable nonclassical features of quantum theory, related to strong correlations between outcomes of measurements performed on quantum systems. Both phenomena can be witnessed by the violation of certain inequalities, the simplest and most important of which are the Clauser-Horne-Shimony-Holt (CHSH) and the Klyachko-Can-Binicio\v{g}lu-Shumovski (KCBS), for Bell nonlocality and Kochen-Specker contextuality, respectively. It has been shown that, using the most common interpretation of Bell scenarios, quantum systems cannot violate both inequalities concomitantly, thus suggesting a monogamous relation between the two phenomena. In this Letter, we show that the joint consideration of the CHSH and KCBS inequalities naturally calls for the so-called generalized Bell scenarios, which, contrary to the previous results, allows for joint violation of them. A photonic experiment thus tests the synchronous violation of both CHSH and KCBS inequalities. Our results agree with the theoretical predictions, thereby providing experimental proof of the coexistence of Bell nonlocality and Kochen-Specker contextuality in the simplest scenario, and shed new light for further explorations of nonclassical features of quantum systems.
\end{abstract}

%%%%%%%%%%%%%%%%%%%%%%%%%%%%%%%%%%%%%%%%%%%%%%%%%%%%%%%%%%%%%%%%%%%

\maketitle

%%%%%%%%%%%%%%%%%%%%%%%%%%%%%%%%%%%%%%%%%%%%%%%%%%%%%%%%%%%%%%%%%%%
% Introduction
%%%%%%%%%%%%%%%%%%%%%%%%%%%%%%%%%%%%%%%%%%%%%%%%%%%%%%%%%%%%%%%%%%%

{\it Introduction.---}
Quantum theory is notable for being intriguing and counter-intuitive, a fact due, mostly, to predictions and concepts that diverge from those of classical theories. Among such nonclassical concepts are Bell nonlocality \cite{Bell_64, BCPSW_14} and Kochen-Specker contextuality \cite{KS_67, Bell_66}.

Classical reasoning assumes the possibility of well-defined values for every physical quantity. Probabilities are used as a consequence of partial knowledge one has regarding \textit{the real state of affairs}. Prior to quantum theory, there was no reason to distrust such a view.

Bell nonlocality refers to stronger-than-classical correlations on outcomes of measurements performed by distant parties on composite systems. Classical reasoning allows for the possibility of the so-called \textit{local hidden variables} (LHV) as a mathematical description of every correlation in space-like separated measurements. In a seminal paper \cite{Bell_64}, Bell showed that quantum theory admits correlations that cannot be explained by any LHV model, a result later known as \textit{Bell's theorem}. The \textit{nonlocal} correlations violate inequalities that are satisfied in any LHV theory, the so-called \textit{Bell inequalities}, the simplest and best known of which is the \textit{Clauser-Horne-Shimony-Holt} (CHSH) inequality \cite{CHSH_69}. It is worth mentioning that Bell nonlocality in quantum systems has been extensively tested and verified in several seminal experiments \cite{FC_72, ADR_82, Hensen_etal_15, Giustina_etal_15, Shalm_etal_15,Weinfurter_etal_16}.

Contextuality is a concept similar to nonlocality; in fact, it can be understood as a generalization of nonlocality that manifests also for single systems. While in classical theories, every pair of measurements can be jointly performed, at least in principle \cite{TerraBundle}, in quantum theory pairs of measurements are usually \textit{incompatible}, preventing their joint measurability. Sets of compatible measurements are called \textit{contexts}, and theories in which it is possible to assign values to the outcomes of measurements irrespective of the context in which they are measured are called \textit{noncontextual hidden variable} (NCHV) theories. Kochen and Specker \cite{KS_67} and Bell \cite{Bell_66} were the first to note that quantum theory admits correlations that cannot be explained by any NCHV model, a result later known as \textit{Kochen--Specker theorem}. In 2008, Klyachko \etal \cite{KCBS_08} noticed that, as for nonlocality, there are inequalities that hold for all noncontextual correlations while can be violated for single quantum systems. The \textit{Klyachko-Can-Binicio\v{g}lu-Shumovski} (KCBS) inequality became the simplest example of contextuality inequality. Interestingly, there are other contextuality inequalities that, in contrast with KCBS, can be violated by \textit{every} quantum state \cite{Cabello_08, YO_12}, thus revealing a property of the measurements known as \textit{state-independent contextuality} (SIC) \cite{Mermin_90,Peres_91,CEG_96}.

Despite their common roots related to the search for hidden variables in quantum theory, nonlocality and contextuality were developed through very distinct research programs, and it was only a few years ago that mathematical approaches to unify both concepts were proposed \cite{CSW_14,AFLS_15, RDTCC_14}. The first proposal to consider nonlocality and contextuality in the same system is due to Kurzy\'{n}ski, Cabello, and Kaszlikowski \cite{KCK_14}: two parties could make a CHSH inequality test while one of the parties would also evaluate the KCBS inequality in a subsystem, using, among others, the same incompatible measurements applied in the nonlocality test.
The authors proved that in quantum theory -- and in more general no-disturbing theories -- there exists a trade-off relation between nonlocality and contextuality indicators, allowing for the violation of only one of these tested inequalities, and conjectured a fundamental \textit{monogamy} relation between nonlocality and (state-dependent) contextuality.
This monogamy relation was experimentally verified by Zhan \etal~\cite{ZZLZSX_16}; and more general monogamy relations were also identified in other scenarios \cite{TV_06, JWG_16, RSKK_12, APTA_14, RH_14, SR_17, PC_09}. It is worth mentioning, though, that such monogamy relations do not hold for state-independent contextuality, since the contextuality test is trivial. Recently, simultaneous observation of Bell nonlocality and state-independent Kochen-Specker contextuality was reported \cite{Hu_etal_18}.

In the work by Kurzy\'{n}ski \etal~\cite{KCK_14}, there was the implicit assumption that nonlocality tests could consider only one measurement from each party. Since, for instance, a test of the KCBS inequality demands five different measurements, each of which performed in a context with a second compatible one, each measurement used in CHSH violation could also be supplemented by a compatible one, leading to new generalized Bell inequalities, in the sense of Ref.~\cite{TRC_19}.

In this Letter, we revisit the scenario considered in Ref.~\cite{KCK_14} and, by considering compatible measurements in the nonlocality test, we experimentally demonstrate that quantum systems can, actually, lead to the synchronous violation of both CHSH and KCBS inequalities. We prove, thus, that there is no fundamental monogamy relation between nonlocality and state-dependent contextuality, even in this simplest scenario where such monogamy was believed to hold. Given that both Bell nonlocality and Kochen-Specker contextuality are important resources for quantum information processing protocols, we believe that this work may be a first step in the direction of devising novel information processing tasks where both nonclassical resources can be used concomitantly.

{\it The scenario:---}
Consider the following measurement scenario (its main ideas and concepts can be extended in a straightforward manner to more general scenarios): two parties, Alice and Bob, run several rounds of experiments on spatially separated laboratories, each on its respective subsystem of a composite physical system, identically prepared in every round. Let Alice be able to perform $m_{A} = |\mathcal{X}|$ possible measurements, labelled by $x \in \mathcal{X}$, each with $o_{A} = |\mathcal{A}|$ possible outcomes, labelled by $a \in \mathcal{A}$. Let Bob be able to perform $m_{B} = |\mathcal{Y}|$ possible measurements, labelled by $y \in \mathcal{Y}$, each with $o_{B} = |\mathcal{B}|$ possible outcomes, labelled by $b \in \mathcal{B}$. Assume, additionally, that some measurements of Bob are \textit{compatible}, meaning that, in each round, Bob is able to perform subsets of measurements concomitantly. Let $\mathcal{C} = \DE{\y}$ be the set \textit{contexts} of Bob, each element $\y$ of which represents a tuple of compatible measurements. Let $\bb \in \mathcal{B}^{|\y|}$ be the (ordered) tuple of outcomes of the tuple of measurements $\y$. After sufficiently many rounds, the parties are able to estimate the following set of probabilities, the so-called \textit{behavior} of the experiment:
\begin{align}
	\p = \DE{p\de{a,\bb|x,\y} \middle| a \in \mathcal{A}, \bb \in \mathcal{B}^{|\y|}, x \in \mathcal{X}, \y \in \mathcal{C}}.
\end{align}

Let the measurements be performed in an informationally separated way, so that the following \textit{no-signalling conditions} hold:
\begin{subequations}
\begin{align}
	\sum_{a}p\de{a,\bb|x,\y} & = p\de{\bb|x,\y} = p\de{\bb|\y}, \; \forall \; \bb,\y,\\
	\sum_{\bb}p\de{a,\bb|x,\y} & = p\de{a|x,\y} = p\de{a|x}, \;\forall \; a,x.
\end{align}
\end{subequations}
Following the formalism presented in \cite{TRC_19}, we define the behavior to be \textit{local} in this scenario, or, in other words, to admit an LHV model, if there are a variable $\lambda$, and probability distributions $p\de{\lambda}$, $p\de{a|x,\lambda}$, and $p\de{\bb|\y,\lambda}$ such that, for all outcomes and measurements:
\begin{align}\label{eq:eLHV}
	p\de{a,\bb|x,\y} = \int p\de{a|x,\lambda} p\de{\bb|\y,\lambda} p\de{\lambda} d\lambda.
\end{align}

Consider the marginal behavior of Bob's experiment:
\begin{align}
	\p_{B} = \DE{p\de{\bb|\y} \middle| \bb \in \mathcal{B}^{\abs{\y}}, \y \in \mathcal{C}}.
\end{align}
Assume it obeys the \textit{no-disturbance} conditions:
\begin{align}
	\sum_{\bb/b} p\de{\bb|\y} = p\de{b|\y} = p\de{b|y}, \; \forall \; b,y,
\end{align}
where $\bb/b$ means that the sum is over all labels in $\bb$ except $b$.
We define the marginal behavior of Bob to be \textit{noncontextual}, or to admit an NCHV model, if there are a variable $\sigma$, and probability distributions $p\de{b|y,\sigma}$, and $p\de{\sigma}$ such that, for all outcomes of all contexts:
\begin{align}
	p\de{\bb|\y} = \int \De{\prod_{y \in \y} p\de{b|y,\sigma}}  p\de{\sigma} d \sigma.
\end{align}

\begin{figure}[htbp]
\includegraphics[width=0.3\textwidth]{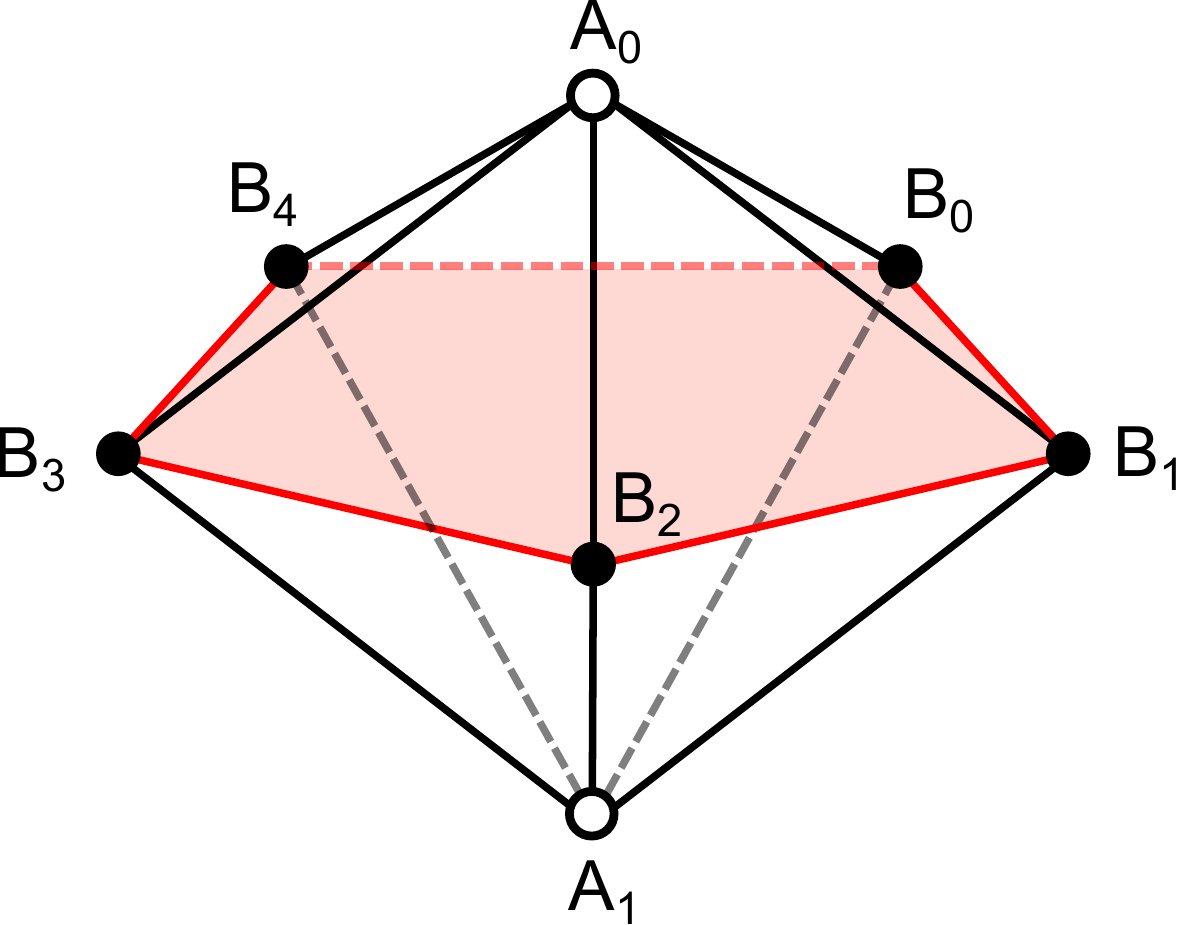}
\caption{Representation of the compatibility relations between all measurements in the experiment. Each measurement is represented by a vertex and vertices connected by an edge represent compatible measurements. Both measurements of Alice (white vertices) are compatible with all measurements of Bob (black vertices); compatibility of measurements of Bob is represented by a pentagon. Sets of measurements that are two-by-two compatible are jointly compatible (\eg:  $\DE{A_{0},B_{0},B_{1}}$).}
\label{fig:1}
\end{figure}

With all these definitions in place, let us focus on the particular scenario we are interested in.
Let Alice choose between two dichotomic measurements, $\mathcal{X} = \DE{0,1}$, $\mathcal{A} = \DE{-1,1}$ and let Bob have five dichotomic measurements, $\mathcal{Y} = \DE{0,1,2,3,4}$, $\mathcal{B} =  \DE{-1,1}$, available, with measurement contexts $\mathcal{C} = \DE{\DE{0,1}, \DE{1,2}, \DE{2,3}, \DE{3,4},\DE{4,0}}$.
The compatibility relations between all measurements are represented in Fig.~\ref{fig:1}. The following version of the CHSH inequality holds for all behaviors that are local (according to the definition in \eqref{eq:eLHV}):
\begin{align}\label{eq:eCHSH}
	\alpha_{\textnormal{CHSH}}=&\mean{A_{0}B_{0}} + \mean{A_{0}B_{2}B_{3}} \nonumber \\
	 &+ \mean{A_{1}B_{0}} - \mean{A_{1}B_{2}B_{3}}\overset{\text{LHV}}{\leqslant}2,
\end{align}
where
\begin{subequations}\label{eq:correlatorsAB}
\begin{align}
	\mean{A_{x}B_{y}}  =& p\de{a=b|x,y} - p\de{a\neq b|x,y}, \\
	\mean{A_{x}B_{y}B_{y'}} = & p\de{a=b\cdot b'|x,y,y'} \nonumber \\
	& - p\de{a\neq b\cdot  b'|x,y,y'}.
\end{align}
\end{subequations}
Note that, according to the definitions above, the joint measurement of $B_{2}B_{3}$ can be regarded as a single dichotomic measurement whose outcome is given by the product $b \cdot b'$, where $b$ and $b'$ are the outcomes of $B_{2}$ and $B_{3}$. The left-hand side of inequality \eqref{eq:eCHSH} is, in essence, equivalent to the left-hand side of the standard CHSH inequality, hence the same local bound.

The marginal scenario of Bob is exactly the one considered by Klyachko and co-authors \cite{KCBS_08}. The marginal behavior $\p_{B}$ is contextual if and only if it violates the KCBS inequality (or one of the inequalities obtained from it by relabelings of measurements and/or outcomes):
\begin{align}\label{eq:KCBS}
	\beta_{\textnormal{KCBS}} =& \mean{B_{0}B_{1}} + \mean{B_{1}B_{2}} + \mean{B_{2}B_{3}} \nonumber \\ & + \mean{B_{3}B_{4}} - \mean{B_{4}B_{0}} \overset{\text{NCHV}}{\leqslant}3,
\end{align}
where
\begin{align}\label{eq:correlatorsBB}
	\mean{B_{y}B_{y'}} & = p\de{b=b'|y,y'} - p\de{b\neq b'|y,y'}.
\end{align}

The main theoretical result of this manuscript is the following:

{\it Theorem:} The CHSH inequality \eqref{eq:eCHSH} and the KCBS inequality \eqref{eq:KCBS} can be concomitantly violated in quantum theory.

{\it Proof:}
We give a direct proof by showing state spaces, measurements and a parameterized family of states obeying all the conditions of the scenario and leading to concomitant violations of both inequalities, for some values of the parameters. The systems of Alice and Bob are a qubit and a qutrit, respectively, whose corresponding basis states are $\{\ket{0},\ket{1}\}$ and $\{\ket{0},\ket{1},\ket{2}\}$.
\begin{itemize}
	\item[(i)] For Alice, choose measurement $x$ to be given by the Pauli observables $A_x$:
	\begin{align}\label{eq:measurements_A}
		A_{0} = \sigma_{z}, \quad A_{1}= \sigma_{x}.
	\end{align}
	
	\item[(ii)] Measurement $y$ of Bob can be represented by an observable $B_{y}$ with eigenvalues in $\DE{\pm 1}$ given as:
	\begin{subequations}\label{eq:measurements_B}
	\begin{align}
		B_{j}= \de{-1}^{j}\de{\one - 2\ketbra{v_{j}}{v_{j}}},
	\end{align}
	for {$j \in \{0,1,2,3,4 \}$}, where $\one$ is the identity matrix and
	\begin{align}
		\ket{v_{j}} \propto & \left[ \cos\de{\frac{4\pi j}{5}}\ket{0} + \sin\de{\frac{4\pi j}{5}}\ket{1} \right. \nonumber \\
		& + \left. \sqrt{\cos\de{\frac{\pi}{5}}}\ket{2} \right].
	\end{align}
	\end{subequations}
	
	Notice that $\braket{v_{j}}{v_{\de{j+1}\!\!\!\mod 5}} = 0$, which implies that $B_{j}$ and $B_{\de{1+j}\!\!\!\mod 5}$ commute, and, hence, are compatible.
	
	\item[(iii)] Consider the one-parameter family of states:
	\begin{subequations}\label{eq:states}
	\begin{align}
		\ket{\Psi\de{\phi}} & = \cos\de{\phi}\ket{u} + \sin\de{\phi}\ket{v},
	\end{align}
	where, for $\theta_{u} \sim 2.868$ and $\theta_{v} \sim 1.449$:
	\begin{align}
		\ket{u} &= \De{\cos\de{\theta_{u}}\ket{0} + \sin\de{\theta_{u}}\ket{1}}\otimes\ket{2}, \\
		\ket{v} &= \De{\cos\de{\theta_{v}}\ket{0} + \sin\de{\theta_{v}}\ket{1}}\otimes\ket{0}.
	\end{align}
	\end{subequations}
\end{itemize}
According to Born's rule, we have:
\begin{subequations}
\begin{align}
	\mean{A_{x}B_{y}B_{y'}} & = \bra{\Psi\de{\phi}} \de{A_{x}\otimes B_{y}B_{y'}}\ket{\Psi\de{\phi}},\\
	\mean{A_{x}B_{y}} & = \bra{\Psi\de{\phi}} \de{A_{x}\otimes B_{y}}\ket{\Psi\de{\phi}}.
\end{align}
\end{subequations}
Any choice of $\phi \in \De{0.288, 0.553}$ completes the proof.
$\square$

In the above proof, the choices of Alice are usual for the maximal CHSH violation and the choices for Bob are usual for the maximal KCBS violation. The parameters $\theta_u$ and $\theta_v$ were obtained by numerical optimization in the corresponding family, for fixed values of $\phi$. Table \ref{tab:1} may help in understanding the role played by $\phi$. This constructive proof allows for a natural experimental verification.

\begin{figure}[htbp]
\includegraphics[width=0.5\textwidth]{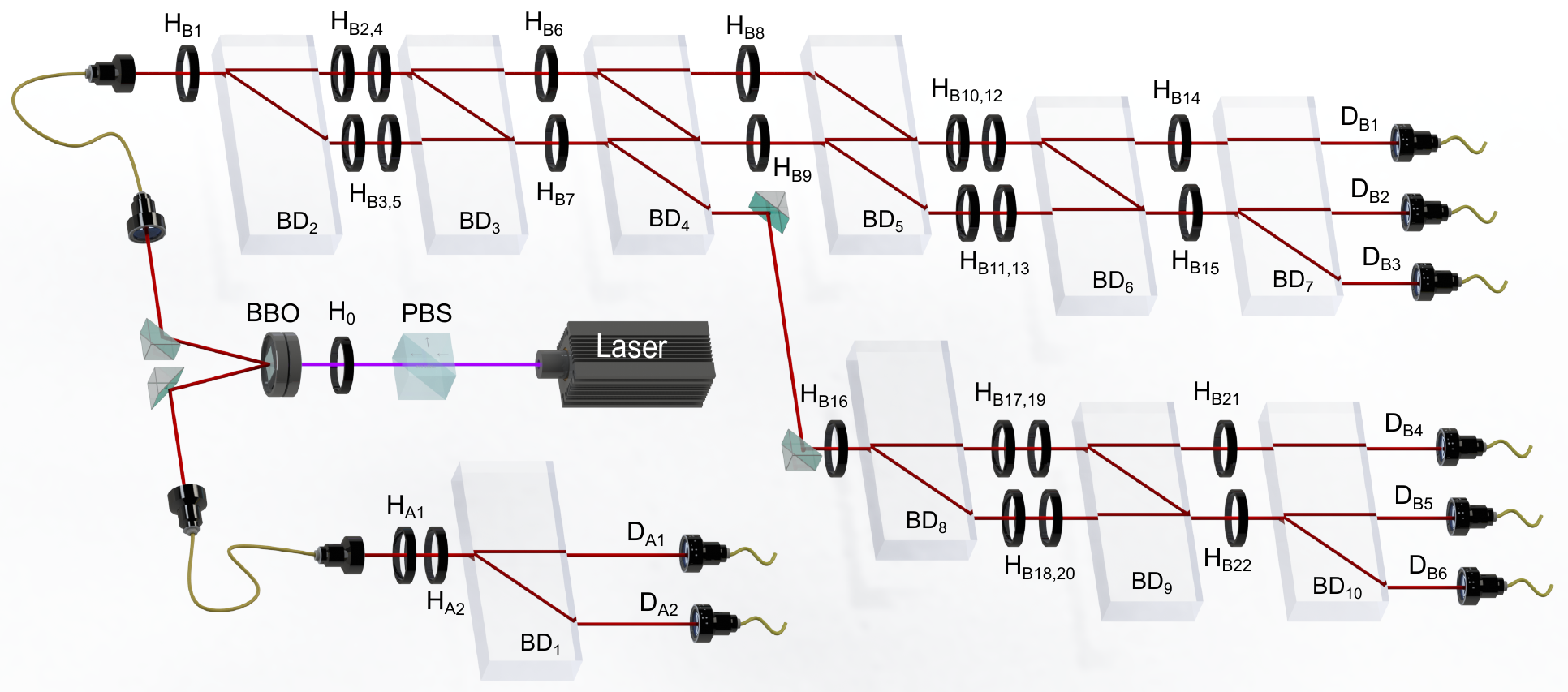}
\caption{Illustration of the experimental setup. Polarization-entangled photon pairs are generated via type-I spontaneous parametric down-conversion where two joint $\beta$-BBO crystals are pumped by a continuous wave diode laser. Qubit is encoded in the horizontal and vertical polarizations of one photon of each pair, while qutrit is encoded in both polarizations and spatial modes of the other photons of the entangled pairs, which are split in different paths dependent on their polarizations via a BD. For Alice, observables $A_i$ are measured via standard polarization measurements using a HWP and a BD. For Bob, cascade Mach-Zehnder interferometers for sequentially measuring observables $B_j$ and $B_{(j+1)\!\!\!\mod 5}$ are used to test the KCBS inequality.}
\label{fig:2}
\end{figure}

\begin{figure}[htbp]
\includegraphics[width=0.4\textwidth]{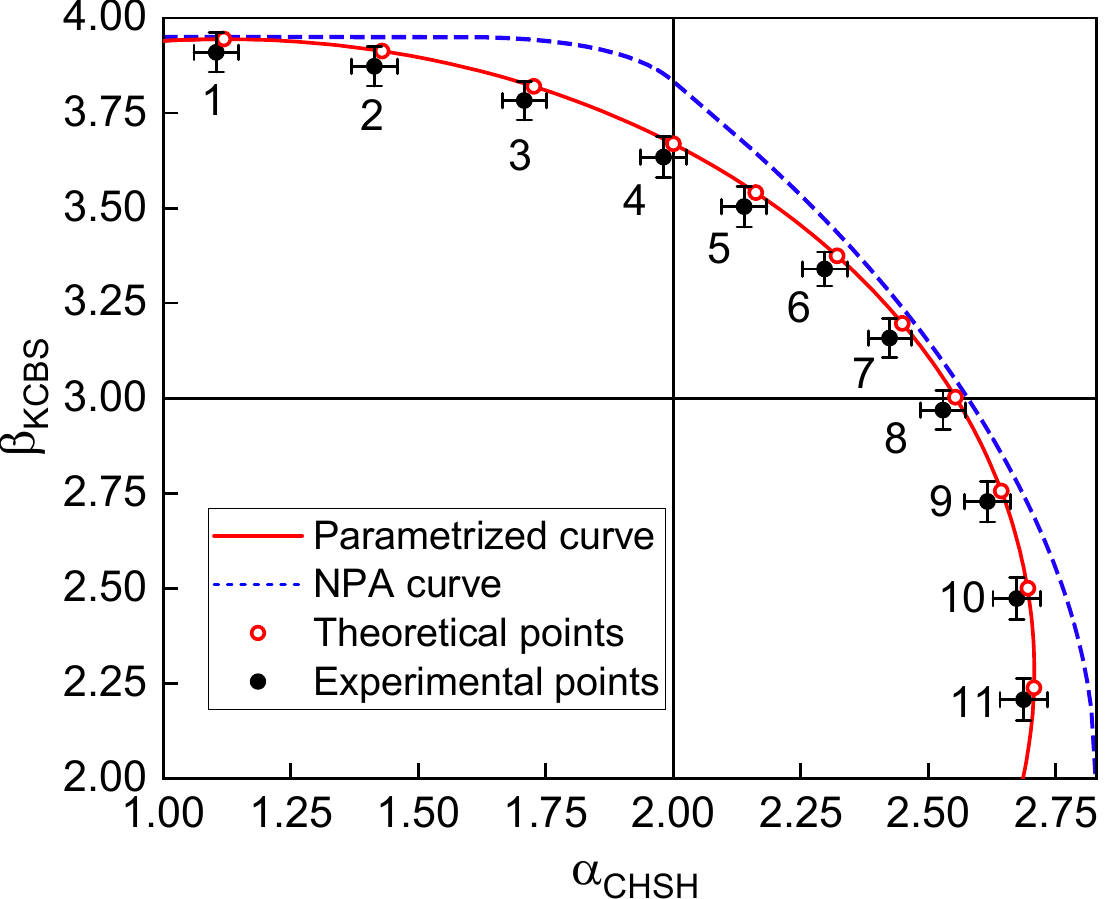}
\caption{Experimental results. The measurements in Eqs. \eqref{eq:measurements_A} and \eqref{eq:measurements_B} and the one-parameter family of states in Eq.~\eqref{eq:states} lead to the solid (red) line. Experimental data of $\alpha_{\textnormal{CHSH}}$ and $\beta_{\textnormal{KCBS}}$ for specific values of the parameter $\phi$ are represented by the black dots and compared to their theoretical predictions (red circles). The points can be separated in three sets: points 1-4 exhibit only contextuality; points 5-7 exhibit both nonlocality and contextuality; and points 8-11 exhibit only nonlocality. Error bars are due to the statistical uncertainty in photon-number counting. Traced (blue) curve is an outer bound to the set of quantum  behaviors calculated by means of the Navascu\'{e}s-Pironio-Ac\'{i}n (NPA) hierarchy \cite{NPA_08}.}
  \label{fig:3}
\end{figure}

\begin{table}[htp]
\centering
\setlength{\tabcolsep}{1mm}{
 \begin{tabular}{c c c c c c}
 \hline
 \hline
State & $\phi$(rad) & $\alpha^\text{th}_{\textnormal{CHSH}}$ & $\alpha^\text{exp}_{\textnormal{CHSH}}$  & $\beta^\text{th}_{\textnormal{KCBS}}$ & $\beta^{\exp}_{\textnormal{KCBS}}$\\
 \hline
$|\Psi_1 \rangle$ & $0$ & $1.1188$ & $1.1043(438)$ & $3.9443$ & $3.9069(518)$\\
$|\Psi_2 \rangle$ & $0.096$ & $1.4293$ & $1.4141(448)$ & $3.9129$ & $3.8728(514)$\\
$|\Psi_3 \rangle$ & $0.192$ & $1.7269$ & $1.7083(438)$ & $3.8199$ & $3.7826(510)$\\
$|\Psi_4 \rangle$ & $0.288$ & $2.0005$ & $1.9813(450)$ & $3.6688$ & $3.6339(536)$\\ \hline
$|\Psi_5 \rangle$ & $0.351$ & $2.1622$ & $2.1382(442)$ & $3.5405$ & $3.5034(529)$\\
$|\Psi_6 \rangle$ & $0.421$ & $2.3215$ & $2.2972(446)$ & $3.3739$ & $3.3397(446)$\\
$|\Psi_7 \rangle$ & $0.487$ & $2.4495$ & $2.4246(423)$ & $3.1964$ & $3.1580(506)$\\ \hline
$|\Psi_8 \rangle$ & $0.553$ & $2.5536$ & $2.5291(435)$ & $3.0021$ & $2.9684(517)$\\
$|\Psi_9 \rangle$ & $0.631$ & $2.6433$ & $2.6164(451)$ & $2.7556$ & $2.7277(537)$\\
$|\Psi_{10} \rangle$ & $0.708$ & $2.6955$ & $2.6739(468)$ & $2.4998$ & $2.4726(555)$\\
$|\Psi_{11} \rangle$ & $0.785$ & $2.7075$ & $2.6871(465)$ & $2.2379$ & $2.2065(553)$\\
 \hline
 \hline
 \end{tabular}}
 \caption{Experimental data of $\alpha_{\textnormal{CHSH}}$ and $\beta_{\textnormal{KCBS}}$ for eleven input states. Error bars are due to the statistical uncertainty in photon-number counting. States 1-4 violate the KCBS inequality but not the CHSH inequality; states 5-7 violate both inequalities; and states 8-11 violate the CHSH inequality only.}
 \label{tab:1}
\end{table}

{\it Experimental realization.---}
To experimentally test the concomitant violations of both KCBS and CHSH inequalities, we set up an experiment where pairs of photons were employed to encode pairs of qubit-qutrit systems. Schematics of the setup are represented in Fig.~\ref{fig:2}.

In each round of the experiment, the photons are prepared in one of the states $\ket{\Psi(\phi)}$ of the one-parameter family defined in Eq.~\eqref{eq:states}. The qubit system is encoded in the polarization degree-of-freedom of one photon of the entangled pair, and the qutrit system is hybridly encoded in both the polarizations and the spatial modes of the other photon of the pair. Measurement of one of the observables given in Eq.~\eqref{eq:measurements_A} is, then, performed in the qubit photon, while sequential measurements of a pair of compatible observables given in Eq.~\eqref{eq:measurements_B} are performed in the qutrit photon. Details of the implementation are provided in the Supplemental Material~\cite{note}.

We produce eleven points $\alpha_{\textnormal{CHSH}}- \beta_{\textnormal{KCBS}}$, corresponding to eleven different input states $\ket{\Psi_i(\phi)}$ ($i=1,\cdots,11$). The experimental results on the average values of the CHSH and KCBS operators are shown in Fig.~\ref{fig:3} and Table~\ref{tab:1}. Synchronous violation of both KCBS and CHSH inequalities are observed for the states $\ket{\Psi_5(\phi)}$, $\ket{\Psi_6(\phi)}$ and $\ket{\Psi_7(\phi)}$ with $\phi=0.351, 0.421, 0.487$, respectively. For $\ket{\Psi_5(\phi)}$, $\alpha_{\textnormal{CHSH}}=2.1382\pm0.0442$ violates the local bound of the inequality by $3$ standard deviations and is in a great agreement with the quantum prediction, $2.1622$. Also, $\beta_{\textnormal{KCBS}}=3.5034\pm0.0529$ violates the noncontextual bound of the KCBS inequality by $9$ standard deviations and is in great agreement with quantum prediction, $3.5405$. For $\ket{\Psi_6(\phi)}$, the CHSH and KCBS inequalities are violated by $6$ and $7$ standard deviations, respectively. For $\ket{\Psi_7(\phi)}$, the violations are by $10$ and $3$ standard deviations, respectively. %Experimental results indicate a successful experimental demonstration and provide strong evidence for synchronous violation of both contextuality and non-locality imposed by quantum theory.

To validate non-disturbance in the data and the compatibility between pairs of observables of Bob, we computed, for each state, the distance $\sum_{j=1}^{5}(p_j-p_j')^{2}$, where $p_j$ is the estimated probability of outcome $b=1$ of the observable $B_{j}$ measured in one context, and $p_j'$ is the corresponding probability of the same observable measured in the other context. As shown in Supplementary Material, the distances for all the states being tested are small enough $(<0.0005)$, which indicates that a very good level of non-disturbance and compatibility between observables holds in our experiment.

{\it Conclusion and discussion:---}
In Ref.~\cite{KCK_14}, the authors showed that the CHSH inequality and the KCBS inequality could not be violated simultaneously by quantum systems.
Their proof uses a usual hypothesis for Bell scenarios, that each part makes one measurement per round. However, since one part is necessarily measuring other compatible observables in order to show contextuality, it is natural to use the locality concept of Ref.~\cite{TRC_19}.

In this Letter, we show that using a more natural notion of locality, the data available in a joint test of CHSH and KCBS can produce joint violation of both inequalities, defying the concept of monogamy between contextuality and nonlocality. We provide examples of states and measurements that lead to such joint violation, and employ a photonic implementation to experimentally demonstrate, for the first time, the synchronous observation of both Bell nonlocality and state-dependent Kochen-Specker contextuality. Our experimental results agree with theoretical predictions, providing a strong evidence that these phenomena are not only no-monogamous, but may be observed and manipulated in simple quantum systems.

From the fundamental point of view, the results presented in this Letter shine new light in the relationship between two of the most important phenomena of the foundations of quantum physics. From the practical point of view, on the other hand, considering the individual importance of Bell nonlocality and Kochen-Specker contextuality to quantum information science, quantum cryptography and quantum computing, we believe that this work may lead to novel possibilities where both concepts could be jointly employed for quantum information processing protocols.

As mentioned in the introduction, there are several results on monogamy between Bell nonlocality and Kochen-Specker contextuality that use the same definition of locality as the one used by Kursy\'nski, Cabello, and Kaszlikowski \cite{KCK_14}. Clearly, such results cannot be directly transposed to the more natural definition used here. Are there, however, measurement scenarios where both definitions agree on whether nonlocality and contextuality are monogamous? This is an important open question that will be investigated in the future work.

%Some relevant open questions come: in Fig.~\ref{fig:3} we see a gap between our explicit realizations of non-monogamous quantum correlations and the NPA bound \cite{NPA_08, W_15}.
%What is the correct way to close this gap?
%Should we go higher in NPA hierarchy, or are there better choices of measurements and states to generate even more significant simultaneous violations than shown here?
%Other point is that many results about monogamy were obtained using the same definition of locality as Kursy\'nski, Cabello, and Kaszlikowski \cite{KCK_14}.
%Clearly, they cannot be directly transposed to the more natural definition used here.
%Can any of them obtain suitable counterparts?

%%%%%%%
%It is important to mention that the set of behaviors that is generated by means of the above states and measurements is not extremal, in the sense that it does not lie in the boundary of the set of quantum behaviors. This can be seem in Fig.~\ref{fig:3}, where, in the bi-dimensional slice of the set of behaviors defined by the values of $\alpha_{\textnormal{CHSH}}$ and $\beta_{\textnormal{KCBS}}$, we plot our parametrized set together with an outer approximation to the set of quantum behaviors, computed via the Navascués-Pironio-Acín (NPA) hierarchy \cite{NPA_08, W_15}. Note that, even though the curves are clearly distinct, they are relatively close in the region of interest, where both inequalities are violated. Behaviors in the boundary of the quantum set may be hard to parametrize and realize in quantum systems, while behaviors in the set we consider are relatively easy to implement.

%%%%%%%

\begin{acknowledgments}
{\it Acknowledgments:---}
R. R. thanks Pawel Kurzy\'{n}ski for fruitful discussions. This work has been supported by the National Natural Science Foundation of China (Grant Nos. 12025401, U1930402, 12088101), the Brazilian National Council for Scientific and Technological Development (CNPq) via the National Institute for Science and Technology on Quantum Information (INCT-IQ) (Grant No. 465469/2014-0), the São Paulo Research Foundation FAPESP (Grant Nos. 2018/07258-7, 2021/01502-6, 2021/10548-0).
\end{acknowledgments}

\newpage

\onecolumngrid

\section{Supplemental Material for ``Synchronous observation of Bell nonlocality and state-dependent Kochen-Specker contextuality''}

\section{Experimental details}

Our experimental setup consists of three modules: state preparation, Alice's measurement, and Bob's measurement.

In the state preparation module, entangled photons of $810$nm wavelength are generated in a type-I spontaneous parametric down-conversion (SPDC) process where two joint $0.5$mm-thick $\beta$-barium-borate ($\beta$-BBO) crystals are pumped by a CW diode laser with $230$mW of power~\cite{EP99,OE09}. The visibility of the entangled photonic state is larger than $97\%$.

One of the generated photons is sent to Alice for her measurement. The qubit is  encoded in the polarization degree of freedom of Alice's photon, i.e., $\{\ket{0}_{A}=\ket{H},\ket{1}_A=\ket{V}\}$, where $\ket{H}$ represents the state of horizontal polarization and $\ket{V}$ represents vertical polarization.

The second photon is then split by a birefringent calcite beam displacer (BD$_2$) into two parallel spatial modes $\ket{U}$ (up) and $\ket{D}$ (down). Both polarization and spatial degrees of freedom of this photon are used, hybridly, to encode a qutrit, whose basis states are associated to the horizontal polarization in the upper mode, the vertical polarization in the upper mode, and the horizontal polarization in the lower mode, respectively, i.e., $\{\ket{0}_B=\ket{UH},\ket{1}_B=\ket{UV},\ket{2}_B=\ket{DH}\}$.

The preparation stage of the setup was designed to prepare states in the following one-parameter family:
\begin{subequations}\label{eq:states_app}
	\begin{align}
		\ket{\Psi\de{\phi}} & = \cos\de{\phi}\ket{u} + \sin\de{\phi}\ket{v},
	\end{align}
	where, for $\theta_{u} \sim 2.868$ and $\theta_{v} \sim 1.449$:
	\begin{align}
		\ket{u} &= \De{\cos\de{\theta_{u}}\ket{0} + \sin\de{\theta_{u}}\ket{1}}\otimes\ket{2}, \\
		\ket{v} &= \De{\cos\de{\theta_{v}}\ket{0} + \sin\de{\theta_{v}}\ket{1}}\otimes\ket{0}.
	\end{align}
	\end{subequations}
The parameter $\phi$ can be adjusted via tuning the setting angles of the half-wave plates (HWPs, H$_0$, H$_{A1}$ and H$_{B1}$-H$_{B3}$). The angles of the HWPs for state preparation are listed in Table~\ref{tab:HWPangles}.

\begin{table}[htbp!]
\centering
%\footnotesize
\setlength{\tabcolsep}{0.7mm}{
 \begin{tabular}{cccccccccccc}
 \hline
 \hline
$\phi$(rad) & $0$ & $0.096$ & $0.192$ & $0.288$ & $0.351$ & $0.421$ & $0.487$ & $0.553$ & $0.631$ & $0.708$ & $0.785$\\
\hline
$H_0(^\circ)$ & $0$ & $2.716$ & $5.429$ & $8.137$ & $9.907$ & $11.865$ & $13.696$ & $15.500$ & $17.555$ & $19.350$ & $20.253$\\

$H_{A1}(^\circ)$ & $127.224$ & $127.185$ & $127.058$ & $126.819$ & $126.574$ & $126.176$ & $125.610$ & $124.716$ & $122.765$ & $118.275$ & $106.972$\\

$H_{B1}(^\circ)$ & $45$ & $44.584$ & $44.136$ & $43.614$ & $43.199$ & $42.624$ & $41.899$ & $40.852$ & $38.727$ & $34.069$ & $22.600$\\
 \hline
 \hline
 \end{tabular}}
 \caption{Setting angles of HWPs for the state preparation. The angles of $H_ {B2}$ and $H_{B3}$ are fixed at $-45^\circ$ and $0^\circ$, respectively, for preparing any initial state. The relation between $H_{A1}$ and $\phi$ is $H_{A1}=0.25\arcsin(0.988\sin2\phi)$.}
 \label{tab:HWPangles}
\end{table}

For the photon which is sent to Alice, the HWP (H$_{A1}$) is used in the preparation of the state $\ket{\Psi(\phi)}$ in Eq.~(\ref{eq:states_app}). The measurement of observable $A_i$ is a standard polarization measurement using HWP (H$_{A2}$) and a BD (BD$_1$). The HWP (H$_{A2}$) at $0^\circ$ $(22.5^\circ)$ is used to map the eigenstates of the observable  $A_0$ ($A_1$) corresponding to the eigenvalue $+1$ into the horizontally polarized state (and, in a complementary fashion, eigenvalue $-1$ into the vertically polarized state). The photons are detected at D$_{A1}$ and D$_{A2}$ right after BD$_1$. The clicks at D$_{A1}$ and D$_{A2}$ correspond to outcomes $-1$ and $1$, respectively.

In the measurement of Bob's observables $B_j$ and their correlators, we use cascade Mach-Zehnder interferometers in three steps~\cite{ARBC_09,AC12}. Each observable $B_j$  has two eigenvalues $\pm1$, one of which is doubly degenerated. To realize the measurement of $B_j$, we use four HWPs (H$_{B4}$-H$_{B7}$) and two BDs (BD$_3$ and BD$_4$). The angles of H$_{B4}$ and H$_{B7}$ are chosen properly -- as displayed in Table~\ref{tab:HWP_measurements_app} -- so that the photons which are in the state $\ket{v_j}$ in Eq.~(12b) of the main text corresponding to the nondegenerated eigenvalue (degenerated eigenvalues) are mapped to the horizontally (vertically) polarized mode after H$_{B7}$. The HWPs H$_{B5}$ and H$_{B6}$ with fixed setting angles are used to fine tune the phase difference between two arms of interferometers. Beam displacer BD$_4$ separates the photons into two different spatial modes corresponding to the two outcomes.

\begin{table}[ht]
\centering
\setlength{\tabcolsep}{1mm}{
 \begin{tabular}{c c c c c c c c c c c}
 \hline
 \hline
Contexts & $H_{A2}(^\circ)$ & $H_{B4}(^\circ)$ & $H_{B7}(^\circ)$ & $H_{B9}(^\circ)$ & $H_{B10}(^\circ)$ & $H_{B12,19}(^\circ)$ & $H_{B15,22}(^\circ)$ & $H_{B16}(^\circ)$ & $H_{B17}(^\circ)$\\
 \hline
 $B_0B_1$ & $0$ & $0$ & $20.958$ & $-60.985$ & $45$ & $-18$ & $-20.985$ & $24.015$ & $45$\\
 $B_1B_2$ & $0$ & $-18$ & $-20.985$ & $-24.015$ & $27$ & $-36$ & $20.985$ & $65.985$ & $27$\\
 $B_2B_3$ & $0$ & $-36$ & $20.985$ & $-65.985$ & $9$ & $36$ & $20.985$ & $24.015$ & $9$\\
 $B_3B_4$ & $0$ & $36$ & $20.985$ & $-65.985$ & $81$ & $18$ & $-20.985$ & $24.015$ & $81$\\
 $B_4B_0$ & $0$ & $18$ & $-20.985$ & $-24.015$ & $63$ & $0$ & $20.985$ & $65.985$ & $63$\\
 $A_0B_0$ & $0$ & $0$ & $20.958$ & $-60.985$ & $45$ & $-18$ & $-20.985$ & $24.015$ & $45$\\
 $A_1B_0$ & $22.5$ & $0$ & $20.958$ & $-60.985$ & $45$ & $-18$ & $-20.985$ & $24.015$ & $45$\\
 $A_0B_2B_3$ & $0$ & $-36$ & $20.985$ & $-65.985$ & $9$ & $36$ & $20.985$ & $24.015$ & $9$\\
 $A_1B_2B_3$ & $22.5$ & $-36$ & $20.985$ & $-65.985$ & $9$ & $36$ & $20.985$ & $24.015$ & $9$\\
 \hline
 \hline
 \end{tabular}}
 \caption{Setting angles of HWPs for measurement devices. For any context, the angles of $H_{B8,B11,B14,B18,B21}$, $H_{B5,B13,B20}$, and $H_ {B6}$ are fixed at $0^\circ$, $45^\circ$ and $90^\circ$, respectively.}
\label{tab:HWP_measurements_app}
\end{table}

After the measurement of the observable $B_{j}$ -- whose results were mapped to two orthogonal polarizations --, the state of the system has to be transformed to the eigenstates of $B_j$ before the sequential measurement of $B_{(j+1)\!\!\!\mod 5}$ on the same photon~\cite{AC12}. The outcomes of $B_j$ are each directed into identical but separated devices, i.e., H$_{B8,B9}$-BD$_5$-H$_{B10,B11}$ for re-creating the eigenstate corresponding to the degenerated eigenvalues, while H$_{B16}$-BD$_8$-H$_{B17,B18}$ for re-creating the eigenstate corresponding to the nondegenerated eigenvalue.

Then, two identical $B_{(j+1)\!\!\!\mod 5}$ measuring devices (one is constructed by H$_{B12}$-H$_{B15}$, BD$_6$-BD$_7$, and another by H$_{B19}$-H$_{B22}$, BD$_9$-BD$_{10}$) are built, each of which is connected to one of the output ports of the measuring device of $B_j$ (each output port corresponds to either degenerated or non-degenerated eigenstate of $B_j$, respectively). The angles of the HWPs for Bob's measurements are listed in Table~\ref{tab:HWP_measurements_app}. The joint probabilities of the outcomes, for each context, are estimated by coincidence count rates at detectors (D$_{B1}$-D$_{B6}$); the precise assignments of each double (or triple) of outcomes to detectors, for each measured context, are shown in Tables~\ref{tab:assignements_two} and \ref{tab:assignements_three}. %By measuring $A_i$ and $A_{i+1}$ together for a number of photons we obtain the average value $\langle A_iA_{i+1}\rangle$---the terms of KCBS inequality.

\begin{table}[h]
\centering
 \begin{tabular}{c c c c c}
 \hline
 \hline
Observable & $p(-1,-1)$ & $p(-1,1)$ & $p(1,-1)$ & $p(1,1)$\\
 \hline
 $B_0B_1$ & $\sum_{m=1,2;n=4,5}C_{Am,Bn}$ & $\sum_{m=1,2}C_{Am,B6}$ & $\sum_{m=1,2;n=1,2}C_{Am,Bn}$ & $\sum_{m=1,2}C_{Am,B3}$\\
 $B_1B_2$ & $\sum_{m=1,2}C_{Am,B1}$ & $\sum_{m=1,2;n=2,3}C_{Am,Bn}$ & $\sum_{m=1,2}C_{Am,B6}$ & $\sum_{m=1,2;n=4,5}C_{Am,Bn}$\\
 $B_2B_3$ & $\sum_{m=1,2;n=4,5}C_{Am,Bn}$ & $\sum_{m=1,2}C_{Am,B6}$ & $\sum_{m=1,2;n=1,2}C_{Am,Bn}$ & $\sum_{m=1,2}C_{Am,B3}$\\
 $B_3B_4$ & $\sum_{m=1,2}C_{Am,B1}$ & $\sum_{m=1,2;n=2,3}C_{Am,Bn}$ & $\sum_{m=1,2}C_{Am,B6}$ & $\sum_{m=1,2;n=4,5}C_{Am,Bn}$\\
 $B_4B_0$ & $\sum_{m=1,2;n=4,5}C_{Am,Bn}$ & $\sum_{m=1,2}C_{Am,B6}$ & $\sum_{m=1,2;n=1,2}C_{Am,Bn}$ & $\sum_{m=1,2}C_{Am,B3}$\\
 $A_0B_0$ & $\sum_{n=4,5,6}C_{A1,Bn}$ & $\sum_{n=1,2,3}C_{A1,Bn}$ & $\sum_{n=4,5,6}C_{A2,Bn}$ & $\sum_{n=1,2,3}C_{A2,Bn}$\\
 $A_1B_0$ & $\sum_{n=4,5,6}C_{A1,Bn}$ & $\sum_{n=1,2,3}C_{A1,Bn}$ & $\sum_{n=4,5,6}C_{A2,Bn}$ & $\sum_{n=1,2,3}C_{A2,Bn}$\\
 \hline
 \hline
 \end{tabular}
 \caption{Each of the joint probabilities
 $p(B_{j}=\pm 1,B_{(j+1)\!\!\!\mod 5} =\pm 1)$ ($j=1,\cdots,5$), and $p(A_i=\pm1,B_{0} = \pm 1)$  ($i=0,1$) is estimated by the sum of the certain coincidence rates $C_{Am,Bn}$ of the APDs $D_{Am}$ and $D_{Bn}$ with $m=1,2$ and $n=1,\cdots,6$, normalized by the total coincidences between the APDs of Alice and Bob.}
 \label{tab:assignements_two}
 \end{table}

 \begin{table}[h]
\centering
 \begin{tabular}{c c c c c c c c c}
 \hline
 \hline
 Observable & $p(-1,-1,-1)$ & $p(-1,-1,1)$ & $p(-1,1,-1)$ & $p(-1,1,1)$ & $p(1,-1,-1)$ & $p(1,-1,1)$ & $p(1,1,-1)$ & $p(1,1,1)$ \\
 \hline
 $A_0B_2B_3$ &
 $\sum_{n=4,5}C_{A1,Bn}$ & $C_{A1,B6}$ & $\sum_{n=1,2}C_{A1,Bn}$ & $C_{A1,B3}$ & $\sum_{n=4,5}C_{A2,Bn}$ & $C_{A2,B6}$ & $\sum_{n=1,2}C_{A2,Bn}$ & $C_{A2,B3}$\\
  $A_{1}B_{2}B_{3}$& $\sum_{n=4,5}C_{A1,Bn}$ & $C_{A1,B6}$ & $\sum_{n=1,2}C_{A1,Bn}$ & $C_{A1,B3}$
& $\sum_{n=4,5}C_{A2,Bn}$ & $C_{A2,B6}$ & $\sum_{n=1,2}C_{A2,Bn}$ & $C_{A2,B3}$\\
 \hline
 \hline
 \end{tabular}
 \caption{Similarly, each of the joint probabilities $P(A_i=\pm1,B_2=\pm1,B_3=\pm1)$  ($i=0,1$) is estimated by the sum of the certain coincidence rates $C_{Am,Bn}$ of the APDs $D_{Am}$ and $D_{Bn}$ with $m=1,2$ and $n=1,\cdots,6$, normalized by the total coincidences between the APDs of Alice and Bob.}
 \label{tab:assignements_three}
\end{table}

For the photon detection, we only register the coincidence rates between the detectors (single-photon avalanche photodiodes with $3$ns time window) of Alice and Bob. For each measurement, we record clicks for $2$s, and the total coincidence counts are about $5500$. To test the KCBS inequality, the correlators $\langle B_jB_{(j+1)\!\!\!\mod 5}\rangle$ are constructed from the measured joint probabilities, $p\de{b,b'|j,(j+1)\!\!\!\mod 5}$, according to Eq.~(10) of the main text.
%$\mean{B_iB_{(i+1)\!\!\!\mod 5}}=P(B_i=1,B_{(i+1)\!\!\!\mod 5}=1)+P(B_i=-1,B_{(i+1)\!\!\!\mod 5}=-1)-P(B_i=1,B_{(i+1)\!\!\!\mod 5}=-1)-P(B_i=-1,B_{(i+1)\!\!\!\mod 5}=1)$.
Similarly, we can evaluate the value of $\alpha_{\textnormal{CHSH}}$ for the CHSH inequality with the correlators $\langle A_i B_{j}\rangle$ and $\langle A_i B_{j}B_{j'}\rangle$, which are constructed from the measured joint probabilities according to
\begin{subequations}\label{eq:correlatorsAB_app}
\begin{align}
	\mean{A_{x}B_{y}}  =& p\de{a=b|x,y} - p\de{a\neq b|x,y}, \\
	\mean{A_{x}B_{y}B_{y'}} = & p\de{a=b\cdot b'|x,y,y'} - p\de{a\neq b\cdot  b'|x,y,y'}.
\end{align}
\end{subequations}
%$P(A_i=\pm1,B_j=\pm1)$ and $P(A_i=\pm1,B_j=\pm1,B_k=\pm1)$, respectively, i.e, $\langle A_i B_{j}\rangle=P(A_i=1,B_j=1)+P(A_i=-1,B_j=-1)-P(A_i=1,B_j=-1)-P(A_i=-1,B_j=1)$, and $\langle A_i B_{j} B_{k}\rangle=P(A_i=1,B_j=1,B_k=1)+P(A_i=1,B_j=-1,B_k=-1)+P(A_i=-1,B_j=-1,B_k=1)+P(A_i=-1,B_j=1,B_k=-1)-P(A_i=1,B_j=1,B_k=-1)-P(A_i=1,B_j=-1,B_k=1)-P(A_i=-1,B_j=1,B_k=1)-P(A_i=-1,B_j=-1,B_k=-1)$.

It is worth mentioning that in the test of the CHSH inequality (7) of the main text, photon loss opens up a detection efficiency loophole in our experiment. A fair-sampling assumption is then taken here, which assumes the events selected out by the photonic coincidence counts is an unbiased representation of the whole sample.

\section{Experimental data}

In this section, we present the observed values of all correlators that are relevant for the tests of the KCBS and CHSH inequalities. Each of the following Tables corresponds to a specific prepared state, discriminated in the caption.

\begin{table}[htbp]
\centering
 \begin{tabular}{c c c c}
 \hline
 \hline
Observable & Theoretical prediction & Experimental value & $\alpha_{\textnormal{CHSH}}$ or $\beta_{\textnormal{KCBS}}$\\
 \hline
 $\langle A_0B_0\rangle$ & $0.0904$ & $0.0904(102)$ & \multirow{4}*{$1.1043(438)$}\\
 $\langle A_1B_0\rangle$ & $-0.0545$ & $-0.0541(112)$ &  \\
 $\langle A_0B_2B_3\rangle$ & $0.6754$ & $0.6679(108)$ &  \\
 $\langle A_1B_2B_3\rangle$ & $-0.4075$ & $-0.4001(116)$ &  \\
 \hline
 $\langle B_0B_1\rangle$ & $0.7889$ & $0.7828(104)$ & \multirow{5}*{$3.9069(518)$} \\
 $\langle B_1B_2\rangle$ & $0.7889$ & $0.7817(109)$ &  \\
 $\langle B_2B_3\rangle$ & $0.7889$ & $0.7801(104)$ &  \\
 $\langle B_3B_4\rangle$ & $0.7889$ & $0.7828(105)$ &  \\
 $\langle B_4B_0\rangle$ & $-0.7889$ & $-0.7823(96)$ &  \\
 \hline
 \hline
 \end{tabular}
 \caption{Measured expectation values $\langle A_iB_j\rangle$, $\langle A_iB_2B_3\rangle$ and $\langle B_jB_{(j+1)\!\!\!\mod 5}\rangle$ for the input state $|\Psi_1(\phi)\rangle$ with $\phi=0$. The distance is $\sum_{j=1}^{5}(p_j-p_j')^{2}=0.000016\pm0.000082$.}
\end{table}

\begin{table}[htbp]
\centering
 \begin{tabular}{c c c c}
 \hline
 \hline
Observable & Theoretical prediction & Experimental value & $\alpha_{\textnormal{CHSH}}$ or $\beta_{\textnormal{KCBS}}$\\
 \hline
 $\langle A_0B_0\rangle$ & $0.1631$ & $0.1626(104)$ & \multirow{4}*{$1.4141(448)$}\\
 $\langle A_1B_0\rangle$ & $0.1210$ & $0.1206(114)$ &  \\
 $\langle A_0B_2B_3\rangle$ & $0.6314$ & $0.6251(112)$ &  \\
 $\langle A_1B_2B_3\rangle$ & $-0.5138$ & $-0.5058(118)$ &  \\
 \hline
 $\langle B_0B_1\rangle$ & $0.7946$ & $0.7859(106)$ & \multirow{5}*{$3.8728(514)$} \\
 $\langle B_1B_2\rangle$ & $0.7659$ & $0.7584(98)$ &  \\
 $\langle B_2B_3\rangle$ & $0.7919$ & $0.7829(106)$ &  \\
 $\langle B_3B_4\rangle$ & $0.7659$ & $0.7581(98)$ &  \\
 $\langle B_4B_0\rangle$ & $-0.7946$ & $-0.7875(106)$ &  \\
 \hline
 \hline
 \end{tabular}
 \caption{Measured expectation values $\langle A_iB_j\rangle$, $\langle A_iB_2B_3\rangle$ and $\langle B_jB_{(j+1)\!\!\!\mod 5}\rangle$ for the input state $|\Psi_2(\phi)\rangle$ with $\phi=0.096$. The distance is $\sum_{j=1}^{5}(p_j-p_j')^{2}=0.000029\pm0.000106$.}
\end{table}

\begin{table}[htbp]
\centering
 \begin{tabular}{c c c c}
 \hline
 \hline
Observable & Theoretical prediction & Experimental value & $\alpha_{\textnormal{CHSH}}$ or $\beta_{\textnormal{KCBS}}$\\
 \hline
 $\langle A_0B_0\rangle$ & $0.2334$ & $0.2313(100)$ & \multirow{4}*{$1.7083(438)$}\\
 $\langle A_1B_0\rangle$ & $0.2906$ & $0.2894(118)$ &  \\
 $\langle A_0B_2B_3\rangle$ & $0.5906$ & $0.5859(108)$ &  \\
 $\langle A_1B_2B_3\rangle$ & $-0.6122$ & $-0.6018(112)$ &  \\
  \hline
 $\langle B_0B_1\rangle$ & $0.8100$ & $0.7936(101)$ & \multirow{5}*{$3.7826(510)$} \\
 $\langle B_1B_2\rangle$ & $0.7261$ & $0.7197(103)$ &  \\
 $\langle B_2B_3\rangle$ & $0.7658$ & $0.7577(102)$ &  \\
 $\langle B_3B_4\rangle$ & $0.7261$ & $0.7180(103)$ &  \\
 $\langle B_4B_0\rangle$ & $-0.8100$ & $-0.7936(101)$ &  \\
 \hline
 \hline
 \end{tabular}
 \caption{Measured expectation values $\langle A_iB_j\rangle$, $\langle A_iB_2B_3\rangle$ and $\langle B_jB_{(j+1)\!\!\!\mod 5}\rangle$ for the input state $|\Psi_3(\phi)\rangle$ with $\phi=0.192$. The distance is $\sum_{j=1}^{5}(p_j-p_j')^{2}=0.000461\pm0.000470$.}
\end{table}

\begin{table}[htbp]
\centering
 \begin{tabular}{c c c c}
 \hline
 \hline
Observable & Theoretical prediction & Experimental value & $\alpha_{\textnormal{CHSH}}$ or $\beta_{\textnormal{KCBS}}$\\
 \hline
 $\langle A_0B_0\rangle$ & $0.2987$ & $0.2958(107)$ & \multirow{4}*{$1.9813(450)$}\\
 $\langle A_1B_0\rangle$ & $0.4481$ & $0.4453(111)$ &  \\
 $\langle A_0B_2B_3\rangle$ & $0.5546$ & $0.5512(115)$ &  \\
 $\langle A_1B_2B_3\rangle$ & $-0.6991$ & $-0.6890(117)$ &  \\
  \hline
 $\langle B_0B_1\rangle$ & $0.8076$ & $0.8005(105)$ & \multirow{5}*{$3.6339(536)$} \\
 $\langle B_1B_2\rangle$ & $0.6710$ & $0.6649(109)$ &  \\
 $\langle B_2B_3\rangle$ & $0.7117$ & $0.7038(108)$ &  \\
 $\langle B_3B_4\rangle$ & $0.6710$ & $0.6653(109)$ &  \\
 $\langle B_4B_0\rangle$ & $-0.8076$ & $-0.7994(105)$ &  \\
 \hline
 \hline
 \end{tabular}
 \caption{Measured expectation values $\langle A_iB_j\rangle$, $\langle A_iB_2B_3\rangle$ and $\langle B_jB_{(j+1)\!\!\!\mod 5}\rangle$ for the input state $|\Psi_4(\phi)\rangle$ with $\phi=0.288$. The distance is $\sum_{j=1}^{5}(p_j-p_j')^{2}=0.000020\pm0.000096$.}
\end{table}

\begin{table}[htbp]
\centering
 \begin{tabular}{c c c c}
 \hline
 \hline
Observable & Theoretical prediction & Experimental value & $\alpha_{\textnormal{CHSH}}$ or $\beta_{\textnormal{KCBS}}$\\
 \hline
 $\langle A_0B_0\rangle$ & $0.3375$ & $0.3332(106)$ & \multirow{4}*{$2.1382(442)$}\\
 $\langle A_1B_0\rangle$ & $0.5420$ & $0.5367(107)$ &  \\
 $\langle A_0B_2B_3\rangle$ & $0.5342$ & $0.5304(115)$ &  \\
 $\langle A_1B_2B_3\rangle$ & $-0.7483$ & $-0.7379(114)$ &  \\
  \hline
 $\langle B_0B_1\rangle$ & $0.8120$ & $0.8042(103)$ & \multirow{5}*{$3.5034(529)$} \\
 $\langle B_1B_2\rangle$ & $0.6275$ & $0.6221(108)$ &  \\
 $\langle B_2B_3\rangle$ & $0.6617$ & $0.6538(107)$ &  \\
 $\langle B_3B_4\rangle$ & $0.6275$ & $0.6195(108)$ &  \\
 $\langle B_4B_0\rangle$ & $-0.8120$ & $-0.8038(103)$ &  \\
 \hline
 \hline
 \end{tabular}
 \caption{Measured expectation values $\langle A_iB_j\rangle$, $\langle A_iB_2B_3\rangle$ and $\langle B_j B_{(j+1)\!\!\!\mod 5}\rangle$ for the input state $|\Psi_5(\phi)\rangle$ with $\phi=0.351$. The distance is $\sum_{j=1}^{5}(p_j-p_j')^2=0.000013\pm0.000079$.}
\end{table}

\begin{table}[htbp]
\centering
 \begin{tabular}{c c c c}
 \hline
 \hline
Observable & Theoretical prediction & Experimental value & $\alpha_{\textnormal{CHSH}}$ or $\beta_{\textnormal{KCBS}}$\\
 \hline
 $\langle A_0B_0\rangle$ & $0.3763$ & $0.3719(108)$ & \multirow{4}*{$2.2972(446)$}\\
 $\langle A_1B_0\rangle$ & $0.6355$ & $0.6304(106)$ &  \\
 $\langle A_0B_2B_3\rangle$ & $0.5151$ & $0.5112(117)$ &  \\
 $\langle A_1B_2B_3\rangle$ & $-0.7946$ & $-0.7837(115)$ &  \\
  \hline
 $\langle B_0B_1\rangle$ & $0.8168$ & $0.8079(103)$ & \multirow{5}*{$3.3397(446)$} \\
 $\langle B_1B_2\rangle$ & $0.5732$ & $0.5691(109)$ &  \\
 $\langle B_2B_3\rangle$ & $0.5940$ & $0.5861(109)$ &  \\
 $\langle B_3B_4\rangle$ & $0.5732$ & $0.5676(109)$ &  \\
 $\langle B_4B_0\rangle$ & $-0.8168$ & $-0.8090(103)$ &  \\
 \hline
 \hline
 \end{tabular}
 \caption{Measured expectation values $\langle A_iB_j\rangle$, $\langle A_iB_2B_3\rangle$ and $\langle B_j B_{(j+1)\!\!\!\mod 5}\rangle$ for the input state $|\Psi_6(\phi)\rangle$ with $\phi=0.421$. The distance is $\sum_{j=1}^{5}(p_j-p_j')^{2}=0.000011\pm0.000071$.}
\end{table}

\begin{table}[htbp]
\centering
 \begin{tabular}{c c c c}
 \hline
 \hline
Observable & Theoretical prediction & Experimental value & $\alpha_{\textnormal{CHSH}}$ or $\beta_{\textnormal{KCBS}}$\\
 \hline
 $\langle A_0B_0\rangle$ & $0.4078$ & $0.4039(104)$ & \multirow{4}*{$2.4246(423)$}\\
 $\langle A_1B_0\rangle$ & $0.7116$ & $0.7050(99)$ &  \\
 $\langle A_0B_2B_3\rangle$ & $0.5007$ & $0.4975(112)$ &  \\
 $\langle A_1B_2B_3\rangle$ & $-0.8294$ & $-0.8182(108)$ &  \\
  \hline
 $\langle B_0B_1\rangle$ & $0.8212$ & $0.8127(97)$ & \multirow{5}*{$3.1580(506)$} \\
 $\langle B_1B_2\rangle$ & $0.5173$ & $0.5104(104)$ &  \\
 $\langle B_2B_3\rangle$ & $0.5194$ & $0.5119(104)$ &  \\
 $\langle B_3B_4\rangle$ & $0.5173$ & $0.5114(104)$ &  \\
 $\langle B_4B_0\rangle$ & $-0.8212$ & $-0.8116(97)$ &  \\
 \hline
 \hline
 \end{tabular}
 \caption{Measured expectation values $\langle A_iB_j\rangle$, $\langle A_iB_2B_3\rangle$ and $\langle B_j B_{(j+1)\!\!\!\mod 5}\rangle$ for the input state $|\Psi_7(\phi)\rangle$ with $\phi=0.487$. The distance is $\sum_{j=1}^{5}(p_j-p_j')^{2}=0.000071\pm0.000172$.}
\end{table}

\begin{table}[htbp]
\centering
 \begin{tabular}{c c c c}
 \hline
 \hline
Observable & Theoretical prediction & Experimental value & $\alpha_{\textnormal{CHSH}}$ or $\beta_{\textnormal{KCBS}}$\\
 \hline
 $\langle A_0B_0\rangle$ & $0.4340$ & $0.4300(108)$ & \multirow{4}*{$2.5291(435)$}\\
 $\langle A_1B_0\rangle$ & $0.7746$ & $0.7684(100)$ &  \\
 $\langle A_0B_2B_3\rangle$ & $0.4901$ & $0.4885(116)$ &  \\
 $\langle A_1B_2B_3\rangle$ & $-0.8549$ & $-0.8422(111)$ &  \\
  \hline
 $\langle B_0B_1\rangle$ & $0.8254$ & $0.8198(99)$ & \multirow{5}*{$2.9684(519)$} \\
 $\langle B_1B_2\rangle$ & $0.4578$ & $0.4518(107)$ &  \\
 $\langle B_2B_3\rangle$ & $0.4359$ & $0.4285(107)$ &  \\
 $\langle B_3B_4\rangle$ & $0.4578$ & $0.4533(107)$ &  \\
 $\langle B_4B_0\rangle$ & $-0.8254$ & $-0.8150(99)$ &  \\
 \hline
 \hline
 \end{tabular}
 \caption{Measured expectation values $\langle A_iB_j\rangle$, $\langle A_iB_2B_3\rangle$ and $\langle B_j B_{(j+1)\!\!\!\mod 5}\rangle$ for the input state $|\Psi_8(\phi)\rangle$ with $\phi=0.553$. The distance is $\sum_{j=1}^{5}(p_j-p_j')^2=0.000012\pm0.000073$.}
\end{table}

\begin{table}[htbp]
\centering
 \begin{tabular}{c c c c}
 \hline
 \hline
Observable & Theoretical prediction & Experimental value & $\alpha_{\textnormal{CHSH}}$ or $\beta_{\textnormal{KCBS}}$\\
 \hline
 $\langle A_0B_0\rangle$ & $0.4572$ & $0.4515(113)$ & \multirow{4}*{$2.5714(451)$}\\
 $\langle A_1B_0\rangle$ & $0.8308$ & $0.8228(102)$ &  \\
 $\langle A_0B_2B_3\rangle$ & $0.4828$ & $0.4830(121)$ &  \\
 $\langle A_1B_2B_3\rangle$ & $-0.8726$ & $-0.8591(115)$ &  \\
  \hline
 $\langle B_0B_1\rangle$ & $0.8299$ & $0.8238(102)$ & \multirow{5}*{$2.7277(537)$} \\
 $\langle B_1B_2\rangle$ & $0.3841$ & $0.3788(111)$ &  \\
 $\langle B_2B_3\rangle$ & $0.3274$ & $0.3228(111)$ &  \\
 $\langle B_3B_4\rangle$ & $0.3841$ & $0.3803(111)$ &  \\
 $\langle B_4B_0\rangle$ & $-0.8299$ & $-0.8220(102)$ &  \\
 \hline
 \hline
 \end{tabular}
 \caption{Measured expectation values $\langle A_iB_j\rangle$, $\langle A_iB_2B_3\rangle$ and $\langle B_j B_{(j+1)\!\!\!\mod 5}\rangle$ for the input state $|\Psi_9(\phi)\rangle$ with $\phi=0.631$. The distance is $\sum_{j=1}^{5}(p_j-p_j')^{2}=0.000048\pm0.000148$.}
\end{table}

\begin{table}[htbp]
\centering
 \begin{tabular}{c c c c}
 \hline
 \hline
Observable & Theoretical prediction & Experimental value & $\alpha_{\textnormal{CHSH}}$ or $\beta_{\textnormal{KCBS}}$\\
 \hline
 $\langle A_0B_0\rangle$ & $0.4657$ & $0.4658(118)$ & \multirow{4}*{$2.6739(468)$}\\
 $\langle A_1B_0\rangle$ & $0.8594$ & $0.8594(105)$ &  \\
 $\langle A_0B_2B_3\rangle$ & $0.4804$ & $0.4805(126)$ &  \\
 $\langle A_1B_2B_3\rangle$ & $-0.8682$ & $-0.8682(119)$ &  \\
  \hline
 $\langle B_0B_1\rangle$ & $0.8250$ & $0.8250(105)$ & \multirow{5}*{$2.4726(555)$} \\
 $\langle B_1B_2\rangle$ & $0.3065$ & $0.3065(115)$ &  \\
 $\langle B_2B_3\rangle$ & $0.2121$ & $0.2121(115)$ &  \\
 $\langle B_3B_4\rangle$ & $0.3056$ & $0.3056(115)$ &  \\
 $\langle B_4B_0\rangle$ & $-0.8233$ & $-0.8234(105)$ &  \\
 \hline
 \hline
 \end{tabular}
 \caption{Measured expectation values $\langle A_iB_j\rangle$, $\langle A_iB_2B_3\rangle$ and $\langle B_j B_{(j+1)\!\!\!\mod 5}\rangle$ for the input state $|\Psi_{10}(\phi)\rangle$ with $\phi=0.708$. The distance is $\sum_{j=1}^{5}(p_j-p_j')^{2}=0.000025\pm0.000100$.}
\end{table}

\begin{table}[htbp]
\centering
 \begin{tabular}{c c c c}
 \hline
 \hline
Observable & Theoretical prediction & Experimental value & $\alpha_{\textnormal{CHSH}}$ or $\beta_{\textnormal{KCBS}}$\\
 \hline
 $\langle A_0B_0\rangle$ & $0.4771$ & $0.4723(117)$ & \multirow{4}*{$2.6871(465)$}\\
 $\langle A_1B_0\rangle$ & $0.8787$ & $0.8751(105)$ &  \\
 $\langle A_0B_2B_3\rangle$ & $0.4853$ & $0.4831(125)$ &  \\
 $\langle A_1B_2B_3\rangle$ & $-0.8665$ & $-0.8566(118)$ &  \\
  \hline
 $\langle B_0B_1\rangle$ & $0.8374$ & $0.8306(105)$ & \multirow{5}*{$2.2065(553)$} \\
 $\langle B_1B_2\rangle$ & $0.2352$ & $0.2280(115)$ &  \\
 $\langle B_2B_3\rangle$ & $0.0927$ & $0.0907(113)$ &  \\
 $\langle B_3B_4\rangle$ & $0.2352$ & $0.2271(115)$ &  \\
 $\langle B_4B_0\rangle$ & $-0.8374$ & $-0.8301(105)$ &  \\
 \hline
 \hline
 \end{tabular}
 \caption{Measured expectation values $\langle A_iB_j\rangle$, $\langle A_iB_2B_3\rangle$ and $\langle B_j B_{(j+1)\!\!\!\mod 5}\rangle$ for the input state $|\Psi_{11}(\phi)\rangle$ with $\phi=0.785$. The distance is $\sum_{j=1}^{5}(p_j-p_j')^{2}=0.000025\pm0.000109$.}
\end{table}

\end{document}